\documentclass[12pt]{revtex4}
\usepackage{graphicx}
\begin{document}

\title{Semiclassical Ehrenfest Paths}
\author{Rafael Liberalquino and Fernando Parisio} 
\address{Departamento de F\'{\i}sica, Universidade Federal de Pernambuco, 50670-901,
Recife, Pernambuco, Brazil\\ Fone: 55-081-32227339\\ email: parisio@df.ufpe.br}


\begin{abstract}
Trajectories are a central concept in our understanding of classical phenomena and
also in rationalizing quantum mechanical effects. 
In this work we provide a way to determine semiclassical paths, approximations to quantum averages in phase space, directly from classical trajectories. We avoid the need of intermediate steps, like particular solutions to the Schroedinger equation or numerical integration in phase space by considering the system to be initially in a coherent state and by assuming that its early dynamics is governed by the Heller semiclassical approximation. Our result is valid for short propagation times only, but gives non-trivial information on the quantum-classical transition.
\end{abstract}
\maketitle

\section{Introduction}
\label{Intro}
Even though wave mechanics does not emcompass the concept of trajectory, this classical notion often appears in the quantum formalism. On  a fundamental level, there are trajectory-based
theories like path-integrals \cite{feynman} and Bohmian mechanics \cite{bohm} that aim to provide interpretations and quantitative rules to deal with general quantum systems. 
Within the orthodox structure of wave mechanics we have
the Ehrenfest theorem, which establishes the phase-space paths followed by $\langle \hat{q} \rangle(t)$ and $\langle \hat{p} \rangle(t)$, providing an important, but deceptive, connection with classical mechanics.  An illustration of the continuing interest in the conceptual/experimental status of trajectories in quantum mechanics is a recent implementation of appropriate combinations of strong and weak measurements to determine average paths of single photons in a two-slit interferometer \cite{kocsis}.

The whole of semiclassics is based on Newtonian paths, the goal being to express
quantum mechanical quantities in terms of real \cite{marcel,ludovic,maia} or complex \cite{xavier,marcus,
ribeiro,parisio,marcel2,levkov,goldfarb,ribeiro2} classical trajectories and their stability properties. 
These approximations are extensively used, e.g., in physical chemistry and also in linking the
distinct structures of classical and quantum mechanics, where a critical point is the understanding 
of the emergence of classical chaos from a linear theory. Usually, in order to calculate semiclassical expressions
$\langle \hat{q} \rangle_{sc}(t)$ and $\langle \hat{p} \rangle_{sc}(t)$, we first need an approximate
solution to the time dependent Schroedinger equation $|\psi(t)\rangle_{sc}$. 
The purpose of this manuscript
is to provide a more direct connection between classical, semiclassical and quantum paths, and to discuss the fundamental differences between them. For example, how quantum non-locality shows up in the semiclassical trajectories (and vanishes as $\hbar \rightarrow 0$).

\section{Preliminary remarks}
Given a particle of mass $\mu$ subjected to a potential $V$, the equations of motion for the expectation values of 
position and momentum in quantum mechanics, provided by the Ehrenfest Theorem, lead to
\begin{equation}
\label{ehrenf}
\mu \frac{{\rm d}^2 }{{\rm d}t^2}\langle \hat{q}\rangle_{\psi}=-\left\langle\frac{{\rm d} \hat{V}}{{\rm d} 
\hat{q}}\right\rangle_{\psi}\;,
\end{equation}
where $\langle ... \rangle_{\psi}=\langle \psi(t)| ... | \psi(t) \rangle$, $| \psi(t) \rangle$ being an
arbitrary solution of the time-dependent Schroedinger equation. To simplify the notation we  suppress 
the subscript $\psi$ hereafter.
The formal similarity between the previous relation and Newton's second law is quite misleading. 
This is partly due to the widely known fact that
\begin{equation}
\nonumber
\frac{{\rm d} V}{{\rm d} q}\left|_{\langle \hat{q} \rangle} \ne \left\langle\frac{{\rm d} \hat{V}}{{\rm d} 
\hat{q}}\right\rangle \right.\;.
\end{equation}
The two quantities in the above relation tend to coincide only in the limiting case of spatially localized wave 
functions. 
Some authors refer to the fulfillment of this last condition along with Eq. (\ref{ehrenf}) as the Ehrenfest Theorem \cite{balentine}.
There is, however, another important difference. While in classical mechanics, $\mu\ddot{x}=-{\rm d}V/{\rm d}x$
is a second order differential equation, relation (\ref{ehrenf}) is an {\it identity}, not an equation. Once the 
quantum dynamics is resolved, $| \psi(t) \rangle$ can be used to calculate the quantities in (\ref{ehrenf}) and to show
that both sides of the equality invariably coincide. As a consequence, if one resorts to any approximate method to
get $| \psi(t)\rangle_{approx} \approx | \psi(t)\rangle$, relation (\ref{ehrenf}) ceases to hold exactly. In this case the 
question arises, which side of (\ref{ehrenf}) better describes the exact quantum evolution of the mean values ? 
In principle we are free to pick either 
\begin{equation}
\label{choice}
\frac{{\rm d}^2 }{{\rm d}t^2}\langle \hat{q}\rangle_{approx} \;\; \mbox{or} \;\; -\frac{1}{\mu}
\left\langle\frac{{\rm d} \hat{V}}{{\rm d} \hat{q}}\right\rangle_{approx}
\end{equation}
as our trial approximate acceleration (or a linear combination of them).

\section{Semiclassical Ehrenfest dynamics}
\label{SED}

\subsection{The ingredients}
In the present manuscript we will be concerned with semiclassical approximations $a_{sc}(t)$, $p_{sc}(t)$, and $q_{sc}(t)$ to the corresponding
exact quantum expected values.
Therefore, our basic ingredients are the classical trajectory [$q_c(t)$ and $p_c(t)$] and its stability properties given by the tangent matrix ${\bf m}$.
Consider the initial conditions $(q_0,p_0)$ leading, after a short time, to the phase-space point $(q_f,p_f)$. If we take the neighboring initial point $(q_0+\delta q_0,p_0+\delta p_0)$, one ends up at $(q_f+\delta q_f,p_f+\delta p_f)$. In the linear approximation ${\bf m}$ is defined by
\[ \left( \begin{array}{c}
\frac{\delta q_f}{b} \\
 \frac{\delta p_f}{c} \\
 \end{array} \right)=
 \left( \begin{array}{cc}
 m_{qq} & m_{qp} \\
m_{pq} & m_{pp}  \\
 \end{array} \right)
 \left( \begin{array}{c}
\frac{\delta q_0}{b} \\
 \frac{\delta p_0}{c} \\
 \end{array} \right)\;,\] 
where $b$ and $c$ are constant parameters, that will be conveniently set, with dimensions of position and momentum, respectively.

In the semiclassical domain it is natural to consider localized states, which uniformly tend to a classical point in phase space as 
$\hbar \rightarrow 0$. For this reason, a canonical coherent state is taken as our initial ket,
the corresponding wave function being given by the minimum uncertainty Gaussian packet 
\begin{equation}
\label{gaussian packet}
\psi_0(x)=\langle x| z_0 \rangle= \frac{1}{\pi^{1/4}b^{1/2}}\;e^{-\frac{(x-q_0)^2}{2b^2}+\frac{i}{\hbar}p_0(x-\frac{q_0}{2})}\;, 
\end{equation}
where $| z_0 \rangle$ is a coherent state with $z_0=(q_0/b+ip_0/c)/\sqrt{2}$. The position and momentum uncertainties are 
$\Delta q=b/\sqrt{2}$ and $\Delta p= c/\sqrt{2}$, with $b\,c=\hbar$. These relations also provide convenient scales $b$ and $c$
used in the definition of the stability matrix. In order to satisfy the minimum uncertainty relation we must have $b \sim \hbar^{1-\gamma}$
and $c \sim \hbar^{\gamma}$ for any $\gamma$. However, for $\gamma \ne 1/2$, either $b/c$ or $c/b$ diverges as $\hbar \rightarrow 0$, leading
to a non-physical infinite squeezing in the classical limit. Thus, we assume from now on that
\begin{equation}
\label{bc}
b \sim \sqrt{\hbar}\;\; \mbox{and}\;\; c \sim \sqrt{\hbar}\;,
\end{equation}
so that the classical limit corresponds to a rescaling, the geometrical nature of the state remaining unchanged.
 
Finally, given the classical path and initial state, we choose the simplest semiclassical approximation to describe the time evolution 
of $|z_0 \rangle$, namely, Heller's ``Thawed Gaussian" \cite{heller}:
\begin{equation}
\label{heller}
\psi_{sc}(x,t)=\frac{\pi^{-1/4}b^{-1/2}}{\sqrt{m_{qq}+im_{qp}}}\; e^{-\frac{\zeta}{2b^2}(x-q_c)^2+\frac{i}{\hbar}[S+p_c(x-q_c)+\frac{1}{2}q_0p_0]}\;,
\end{equation}
where $q_c$ and $p_c$ are the classical position and momentum as functions of time, $S$ is the classical action, and $\zeta=(m_{pp}-im_{pq})/(m_{qq}+im_{qp})$. Due to its extreme simplicity, $|\psi_{sc}|^2$ is a Gaussian for all times and the Heller wave function is usually a poor approximation to describe the quantum wave function for longer propagation times. Therefore, our results will be valid only on a short-time scale. This limitation, however, will not prevent us to get useful information.

\subsection{Semiclassical trajectories}
Since the above formula is a Gaussian for all times, its use together with the first option in (\ref{choice}) leads simply to $a_{sc}=a_c$.
In this case, no vestige of the quantum behavior is left. On the contrary, if one employs (\ref{heller}) and the second option in (\ref{choice}),
the result is non-trivial.
Thus, we define the semiclassical acceleration as
\begin{equation}
\label{asc}
a_{sc}\equiv-\frac{1}{\mu}
\left\langle\frac{{\rm d} \hat{V}}{{\rm d} \hat{q}}\right\rangle_{sc}=-\frac{1}{\mu}\int_{-\infty}^{\infty}{\rm d}x |\psi_{sc}(x,t)|^2 \frac{{\rm d}V}{{\rm d}x}\;,
\end{equation}
where $\psi_{sc}(x) =\langle x |\psi(t) \rangle_{sc}$. This definition provides corrections to the bare classical trajectory because the Gaussian extends over the space ``sensing" the different values taken by ${\rm d}V/{\rm d}x$, i. e., the whole force field. It is important to note that the potential cannot be singular
for any finite $x$, since the Gaussian is non-zero everywhere.
We begin our analysis of (\ref{asc}) in a very general way, by assuming that the potential is an arbitrary polynomial:
\begin{equation}
\label{pot}
V(x)=\sum_{n=0}^N\alpha_n x^n\;.
\end{equation}
In principle, $N$ is supposed to be finite, but we may take $N \rightarrow \infty$ in cases where convergence of $a_{sc}$ can be guaranteed.
Substituting the above relation in the general prescription (\ref{asc}) we get
\begin{equation}
\nonumber
a_{sc}=-\frac{1}{\mu}\sum_n\frac{n\alpha_n}{\sqrt{\pi}\sigma(t)}\int_{-\infty}^{\infty}{\rm d}x\,x^{n-1}e^{-\frac{1}{\sigma(t)^2}(x-q_c)^2}\;,
\end{equation}
with $\sigma (t)=b\sqrt{m_{qq}^2+m_{qp}^2}$. This leads to
\begin{equation}
\label{asc2}
a_{sc}(t)=-\frac{1}{\mu}\sum_{n=1}^N n\alpha_n\left[\frac{\sigma(t)}{2i}\right]^{n-1}{\rm H}_{n-1}\left[\frac{iq_c(t)}{\sigma(t)}\right]\;,
\end{equation}
where H$_k$ denotes the Hermite polynomial of degree $k$ [notice that $i^{-k}{\rm H}_k(iy)$ is a real function] and the time dependence was
made explicit. By integration we get $p_{sc}(t)$ and $q_{sc}(t)$. We recall that in order to get these approximations to the quantum
expected values one has only to solve the classical equations for $q_c(t)$ and the stability matrix ${\bf m}$, not the Schroedinger equation. 

\subsection{General properties}
As any consistent semiclassical approximation, $a_{sc}$ must coincide with the classical and exact quantum values for Gaussian packets in linear
and harmonic potentials ($V=const$, $V\propto x$, and $V \propto x^2$). This is indeed the case. For $V=const$, formula (\ref{asc}) gives immediately
$a_{sc}=0$, while for $V=\alpha_1x$, relation (\ref{asc2}) gives the constant acceleration $a_{sc}=-\alpha_1/\mu$. Finally, for $V=\alpha_2x^2$, we get $a_{sc}=-2\alpha_2q_c(t)/\mu$. Setting $\alpha_2=\mu \omega^2/2$ we obtain $a_{sc}(t)=-\omega^2q_c(t)$, as expected. For higher order potentials the classical, semiclassical, and quantum results no longer coincide. 

Before going into more quantitative examples, we establish the general lower order correction to the classical acceleration in the general case of an analytic potential. 
To do that we must realize that the Planck constant appears in (\ref{asc}) [or (\ref{asc2})] implicitly, through $\sigma\propto b \propto \sqrt{\hbar}$. Replacing
\begin{equation}
{\rm H}_n(y)=\sum^{[n/2]}_{k=0}(2y)^{n-2k}(-1)^k\frac{n!}{k!(n-2k)!}
\end{equation}
into (\ref{asc2}) and collecting the lower powers of $\sigma$ we get
\begin{equation}
\label{asc_series}
a_{sc}=a_c-\frac{\sigma^2}{4\mu}\frac{{\rm d}^3 V}{{\rm d q^3}}\left|_{q=q_c} +O(\sigma^4)\right.\;.
\end{equation}
This shows that our result is
consistent with classical mechanics plus a main correction of order $\hbar^1$ and smaller terms (involving higher powers of $\hbar$ and higher derivatives of $V$). Perhaps the most important point about the above relation is that these approximate quantum paths cannot come from any effective or modified potential, because they are not Hamiltonian.
Differently from what happens in classical mechanics, the instantaneous dynamics of a phase-space point depends upon higher order derivatives of the potential. This is a  trace of quantum non-locality. While in the Newtonian mechanics the particle only needs to ``know" its position [to get $V(x)$] and the force to which it is subjected ($\propto$d$V/$d$x$), both local quantities, in the semiclassical case all derivatives of the potential are needed, which amounts to  knowing the function $V$ for all values of $x$, no matter its distance to the particle. 
Any change in the potential in a far region instantly influences the semiclassical particle's path. Of course, this non-locality vanishes as $\hbar \rightarrow 0$.

Note that the correction to the classical acceleration is, in general, non-vanishing even for $t =0$. This can be understood by realizing that our expression yields, by construction, the exact quantum result as $t \rightarrow 0$. For sufficiently short times the evolution operator can be written as $\hat{I}-\frac{i}{\hbar}\hat{H}t-\frac{1}{2\hbar^2}\hat{H}^2t^2$, so that the expectation value of momentum is, to second order in $t$
\begin{eqnarray}
\nonumber
\langle \hat{p} \rangle (t)=\langle z_0 |\left(\hat{I}+\frac{i}{\hbar}\hat{H}t-\frac{1}{2\hbar^2}\hat{H}^2t^2 \right)\hat{p}\left(\hat{I}-\frac{i}{\hbar}\hat{H}t-\frac{1}{2\hbar^2}\hat{H}^2t^2 \right)| z_0 \rangle \\
\nonumber = p_0 +\frac{i}{\hbar} t \langle [\hat{V}(\hat{q}),\hat{p}]\rangle_{z_0} - \frac{t^2}{4 \mu}\left\langle \left\{\hat{p}, \frac{\partial^2 \hat{V}}{\partial \hat{q}^2}\right\} \right\rangle_{z_0} + O(t^3) \\
= p_0-t \left\langle \frac{\partial \hat{V}}{\partial \hat{q}} \right\rangle_{z_0} - \frac{t^2}{4 \mu}\left\langle \left\{\hat{p}, \frac{\partial^2 \hat{V}}{\partial \hat{q}^2}\right\} \right\rangle_{z_0} + O(t^3) \;,
\end{eqnarray}
where $\{\, ,\,\}$ stands for the anticommutator. In the quantum mechanical case (where the Ehrenfest Theorem holds) we define the exact acceleration as
\begin{equation}
a_{quant}=\frac{1}{\mu}\frac{{\rm d}}{{\rm d}t}\langle \hat{p} \rangle (t)=- \frac{1}{\mu}\left\langle \frac{\partial \hat{V}}{\partial \hat{q}} \right\rangle_{z_0}- \frac{t}{2 \mu^2}\left\langle \left\{\hat{p}, \frac{\partial^2 \hat{V}}{\partial \hat{q}^2}\right\} \right\rangle_{z_0} + O(t^2) \;,
\end{equation}
which is $a_{sc}$ for $t=0$. This guarantees a consistent quantum-classical interplay in the sense that the semiclassical expression tends to become classical for $\hbar \rightarrow 0$ while it is the exact quantum result for $t \rightarrow 0$. 
This closes our discussion for short-time propagation in 1D analytic potentials. 
In the next section we illustrate the procedure by applying (\ref{asc2}) in the case of a cubic potential. In addition, we give an example involving a non-analytical potential, for which expression (\ref{asc2}) is not valid and (\ref{asc}) must be used directly. 
\section{Applications}
\label{A}

\subsection{Cubic potential}
Here we consider in more detail the classical, semiclassical, and quantum accelerations in the particular case of a cubic potential $V=\alpha x^3$.  The quantum result up to first order in $t$ is
\begin{equation}
a_{quant}\approx-\frac{3 \alpha}{\mu}\left[ \langle \hat{q}^2\rangle+\frac{t}{\mu}(\langle \hat{q}\hat{p}\rangle + \langle \hat{p}\hat{q}\rangle) \right]= -\frac{3 \alpha}{\mu}
\left[ q_0^2+\frac{b^2}{2}+\frac{2t}{\mu}q_0p_0\right]\;.
\end{equation}
Let us calculate $a_{sc}$ also up to first order in $t$. The full semiclassical expression reads
\begin{equation}
\label{cubic}
a_{sc}=-\frac{3\alpha}{\mu}\left[ q_c(t)^2+\frac{\sigma(t)^2}{2}\right]\;
\end{equation}
which coincides with (\ref{asc_series}) for a cubic potential. We have
$q_c(t)=q_0+\left.\frac{{\rm d} q}{{\rm d} t} \right|_0t+ O(t^2)$,
from which it is easy to show that $m_{qq} \sim 1$, $m_{qp} \sim t^2$, so that $\sigma^2 = b^2+O(t^2)$. Substituting these expansions into (\ref{cubic}) we get 
\begin{equation}
a_{sc}\approx -\frac{3 \alpha}{\mu}
\left[ q_0^2+\frac{b^2}{2}+\frac{2t}{\mu}q_0p_0\right]\;.
\end{equation}
Thus, in this case, the quantum and semiclassical results coincide not only to zeroth order, but also to first order in $t$. On the other hand we know from (\ref{asc_series}) that $a_{sc}$ goes to $a_c$ as $\hbar \rightarrow 0$. We can immediately write
\begin{equation}
a_{c}\approx -\frac{3 \alpha}{\mu}
\left[ q_0^2+\frac{2t}{\mu}q_0p_0\right]\;,
\end{equation}
since $b^2 \sim \hbar$. In particular, it is interesting to note that the classical equilibrium point $q_0=0$ and $p_0=0$, leading to $a_c=0$, does not occur in the semiclassical and quantum cases, since there is always a ``residual" acceleration $a_{sc}=a_{quant}=3 \alpha b^2/2 \mu$.
\subsection{Wave packet impinging on a step potential}
A crucial process in wave-packet dynamics is the collision with a potential wall. Here we examine the evolution of $\langle \hat{q} \rangle$ for the Gaussian packet (\ref{gaussian packet}) bouncing off a step potential,
\begin{equation}
\label{wall}
V(x)=
 \left\{ \begin{array}{c}
0\;, \;\; \mbox{for}\;\; x\le0 \\
 V_{0} \;, \;\; \mbox{for}\;\; x>0\;,\\
 \end{array} \right.
\end{equation}
for the cases where $p_{0}<\sqrt{2\mu V_0}$. 
From the classical trajectory,
\begin{equation}
\label{trajectory1}
q_c(t)=
 \left\{ \begin{array}{c}
\frac{p_0 t}{\mu }+q_0\;, \;\; \mbox{for}\;\; t\le t_0 \\
 -\frac{p_0}{\mu }( t -t_0) \;, \;\; \mbox{for}\;\; t>t_0\;,\\
 \end{array} \right.
\end{equation}
with $t_0\equiv \mu  q_0/p_0$, we extract the two needed tangent matrix elements:
\begin{equation}
\label{tangent matrix 1}
m_{qq}(t)=
 \left\{ \begin{array}{c}
1\;, \;\; \mbox{for}\;\; t\le t_0 \\
-1 \;, \;\; \mbox{for}\;\; t>t_0\;\\
 \end{array} \right.
\end{equation}
and $m_{qp}(t)=\hbar t m_{qq}(t)/\mu b^2 $. 
The semiclassical acceleration, as defined in (\ref{asc}), is simply
\begin{equation}
\label{asc3}
a_{sc}=-\frac{V_0}{\mu}|\psi_{sc}(0,t)|^2
\end{equation}
or, using (\ref{heller}),  (\ref{trajectory1}) and (\ref{tangent matrix 1}),
\begin{equation}
\label{asc4}
a_{sc}(t)=-\frac{V_0}{\mu\sqrt{\pi } \sigma (t)}
 \left\{ \begin{array}{c}
 e^{-\left(\frac{\frac{p t}{\mu}- x_0}{ \sigma (t)}\right)^2}\;, \;\; \mbox{for}\;\; t\le t_0 \\
 e^{-\left(\frac{p(t-t_0)}{\mu \sigma (t)}\right)^2}\;, \;\; \mbox{for}\;\; t>t_0\;,\\
 \end{array} \right.
\end{equation}
where $\sigma (t)=b \sqrt{\left(\frac{\hbar t}{b^2 m}\right)^2+1}$. Numerical integration leads to $q_{sc}(t)$, which is depicted in figure 1 along with $q_c(t)$ and the time evolution of the exact quantum average $\langle \hat{q} \rangle(t)$.  The initial position and momentum are $q_0=0$ and $p_0=1$, and the barrier is located at $q=1$ with height $V=5$, in a system of arbitrary units. The other parameters are $b=0.1$ and  $\mu=1$. In the top panel $\hbar=0.05$ and in the bottom panel $\hbar=0.1$ (Note that the classical action for these trajectories is of order of 1). The time interval in the horizontal axis ends shortly after the classical turning time.
We conclude that even for this discontinuous potential the semiclassical results are quite consistent for times shorter than the classical turning time. In particular, the tendency of the quantum turning point to recede is seen in the semiclassical plots. In fact, this effect may be artificially enhanced in our approximation because in the quantum dynamics the packet deforms such that its penetration into the barrier is smaller in comparison that of the semiclassical Gaussian packet. The deformation is stronger for larger values of $\hbar$ (see the bottom plot).

\section{Conclusion and perspectives}
\label{Concl}

We have developed a direct way to determine semiclassical trajectories for short propagation times. Despite this limitation we were able to reproduce some important quantum features, like the tendency of turning points to recede in the quantum regime. In addition, we identified how quantum non-locality manifests itself in the semiclassical phase-space, destroying the original Hamiltonian structure. In this context it is not possible to resort to any extended Hamiltonian formalism \cite{patt} (see also\cite{cooper}) .

Perhaps, a promising perspective to be drawn from this approach is related to its generalization to higher dimensional problems in connection with chaos. On the one hand we have classical mechanics which presents sensitivity to initial conditions in a variety of systems, and, on the other hand is quantum mechanics, whose linearity prevents exponential separation of $\langle \bf{r} \rangle$ and $\langle \bf{r + {\rm d}r} \rangle$. In principle, our semiclassical paths should be in the middle-way, which raises the question on the behavior of the Liapunov exponents associated to semiclassical paths and on their dependence on $\hbar$.

\section{aknowledgements}
Funding from CNPq and FACEPE (APQ-1415-1.05/10) is gratefully acknowledged.

\begin{figure}
\label{figure1}
\includegraphics[width=10cm]{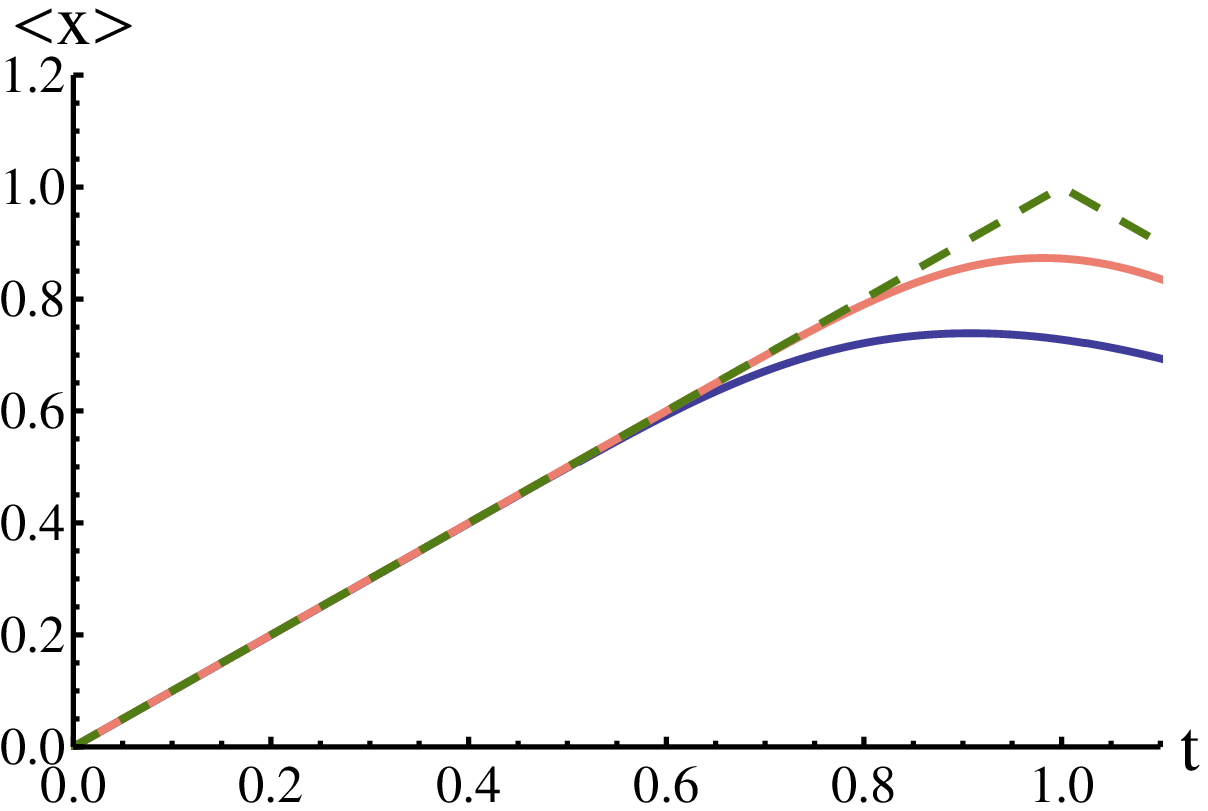}\\
\includegraphics[width=10cm]{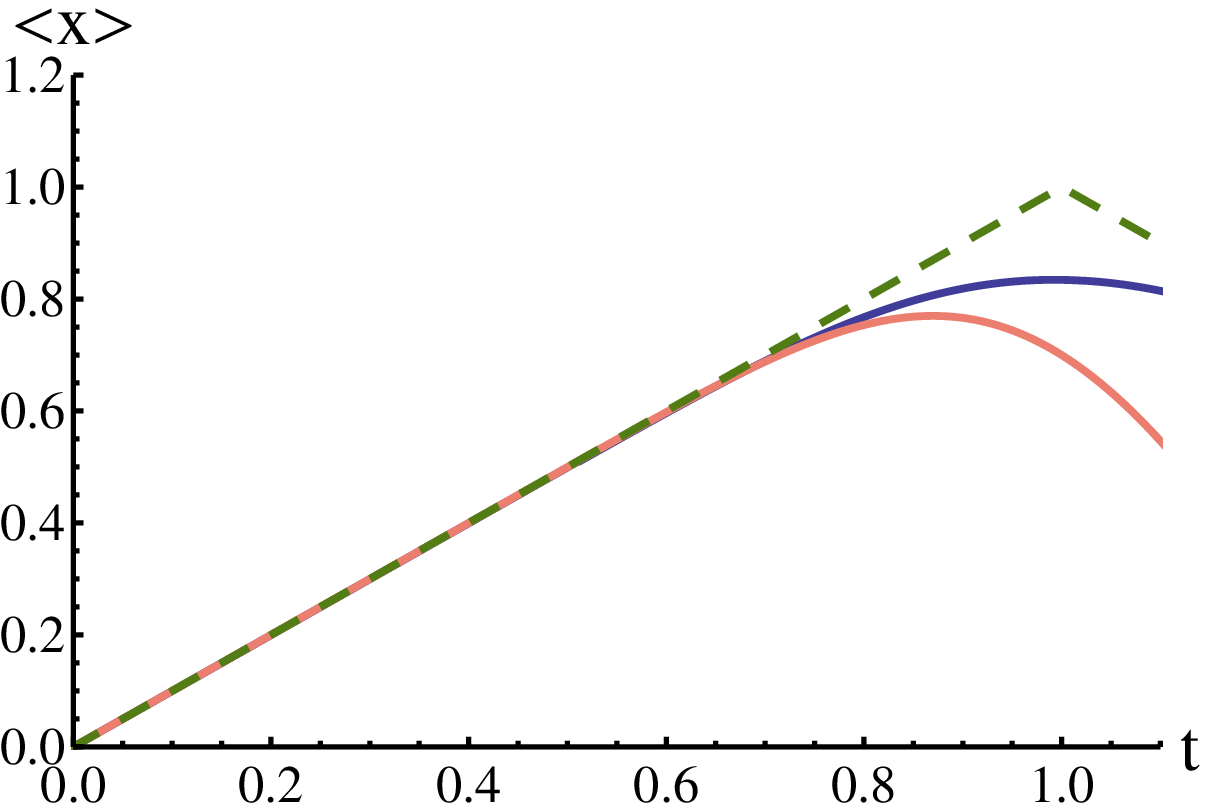}
\caption{Semiclassical trajectories $x_{sc}(t)$ (red), quantum averages $\langle x \rangle (t)$ (blue) and classical paths (green dashed lines). The initial position and momentum are $q_0=0$ and $p_0=1$, the other parameters being $b=0.1$ and $\mu=1$ in a system of arbitrary units. In the left figure $\hbar=0.05$ and in the right figure $\hbar=0.1$. The time interval in the horizontal axis ends soon after the classical turning time. }
\end{figure}
\end{document}